\documentclass[article,pre,tightenlines,notitlepage,showpacs,nofootinbib]{revtex4}
\usepackage{amssymb}
\usepackage{amsmath}
\usepackage{amsfonts}
\usepackage{graphicx}

\input{tcilatex}

\begin{document}

\title[Storage Capacity]{Bistable Gradient Networks II: \ Storage Capacity
and Behaviour Near Saturation}
\author{Patrick McGraw and Michael Menzinger}
\affiliation{University of Toronto }

\begin{abstract}
We examine numerically the storage capacity and the behaviour near
saturation of an attractor neural network consisting of bistable elements
with an adjustable coupling strength, the Bistable Gradient Network (BGN). \
For strong coupling, we find evidence of a first-order \textquotedblleft
memory blackout\textquotedblright\ phase transition as in the Hopfield
network. For weak coupling, on the other hand, there is no evidence of such
a transition and memorized patterns can be stable even at high levels of
loading. \ The enhanced storage capacity comes, however, at the cost of
imperfect retrieval of the patterns from corrupted versions. \ 
\end{abstract}

\date{\today }
\maketitle

\section{ Introduction}

In this paper we consider the behaviour at high memory loading of a
Hopfield-like attractor network of $\ N$ bistable elements, the bistable
gradient network or BGN\cite{Chinarov}\cite{IWANN}. \ This is a sequel to an
earlier paper \cite{LowLoading} which considered the BGN at low loading. \
We compare the BGN to the deterministic Hopfield network (HN)\cite{Hopfield}-%
\cite{Amit2}, examining the storage capacity and other key properties.

To begin, we review the BGN model and establish some notation. \ The BGN is
described by the coupled differential equations 
\begin{equation}
\frac{dx_{i}}{dt}=-\frac{\partial H}{\partial x_{i}},
\end{equation}
where $x_{i}$ are $N$ \ continuous-valued real variables (or components of
an $N$-dimensional state vector $\mathbf{x}$) representing the outputs of
the $N$ nodes of the network, and $H$ is the Hamiltonian 
\begin{equation}
H=H_{0}+H_{int}=\sum_{i=1}^{N}\left( \frac{x_{i}^{4}}{4}-\frac{x_{i}^{2}}{2}%
\right) -\frac{1}{2}\gamma\sum_{i,j=1}^{N}w_{ij}x_{i}x_{j}.
\label{Hamiltonian}
\end{equation}
The quantities $w_{ij}$ are the elements of a symmetric matrix of connection
strengths and $\gamma$ is a control parameter determining the overall
strength of the coupling among nodes. \ As in the Hopfield model\cite%
{Hopfield}\cite{Little}, the connections or synaptic weights are determined
by the Hebb learning rule\cite{Hebb} 
\begin{equation}
w_{ij}=\frac{1}{N}\sum_{\mu=1}^{p}\xi_{i}^{\mu}\xi_{j}^{\mu}-\frac{p}{N}%
\delta_{ij}  \label{Hebb}
\end{equation}
where the N-dimensional vectors $\mathbf{\xi}_{{}}^{\mu}$ represent a set of 
$p$ distinct \emph{memory patterns} to be recognized by the network. \ We
take these patterns to consist of binary elements $\pm1$ only, \ and we
assign them random values, thus introducing quenched disorder. \ The BGN's
key difference from the HN and from most of its continuous-valued relatives 
\cite{ContinuousHop} lies in the presence of the local quartic potential
term $H_{0}$ in the Hamiltonian, which renders each node bistable. \ 

The interaction term $H_{int}$ furnishes an input to each node given by $%
h_{i}=\gamma\sum_{\mu=1}^{p}w_{ij}x_{j}$ so that the dynamical equation for
each node is given by 
\begin{equation}
\frac{dx_{i}}{dt}=x_{i}-x_{i}^{3}+h_{i}.
\end{equation}
The input may also be referred to as a \textquotedblleft magnetic
field\textquotedblright\ by analogy with Ising spin systems. \ In the
absence of input each node has two stable fixed points at $\pm1$, but a
nonzero field shifts the two fixed points. \ If the critical magnitude $%
h_{c}=\frac {2\sqrt{3}}{9}\approx0.38$ is exceeded then the field is strong
enough to overcome the potential barrier in the quartic potential and there
is then only one fixed point.

We define two sets of order parameters $m_{\mu}$ and $b_{\mu}$ by 
\begin{equation}
m_{\mu}\equiv\mathbf{\xi}^{\mu}\cdot\mathbf{x}=\frac{1}{N}%
\sum_{i=0}^{N}\xi_{i}^{\mu}x_{i}  \label{mdef}
\end{equation}
and%
\begin{equation}
b_{\mu}\equiv\frac{1}{N}\sum_{i=0}^{N}\xi_{i}^{\mu}\func{sign}(x_{i}).
\label{bdef}
\end{equation}
$m_{\mu}$ are inner products or overlaps of the state vector with the
memorized patterns, while $b_{\mu}$, the \textquotedblleft bit
overlaps,\textquotedblright\ encode information about sign agreements
between the state vector and the stored patterns. \ For the purposes of this
paper, we will for the most part be more interested in $b_{\mu}$ than in $%
m_{\mu}$, so where there is little risk of confusion we will drop the word
\textquotedblleft bit\textquotedblright\ and simply refer to $b_{\mu}$ as an
\textquotedblleft overlap.\textquotedblright

The degree of loading of the network's memory can be parametrized by the
ratio $\alpha\equiv p/N$. \ \ In the companion paper \cite{LowLoading} we
examined the behaviour of the BGN in the low-loading limit $\alpha\ll1$. \ \
It was shown that the network can function as an associative memory and
correct sign flip errors in a stored pattern as long as $\gamma>\frac{1}{3}$%
\ . \ \ We found that the attractors of the BGN's dynamics are readily
classified into three categories which are separated from each other in
energy. \ The lowest energy states are the \emph{memory }or \emph{retrieval
states}, \ each of which corresponds to one of the memorized patterns. \ In
addition to these there are higher-energy spurious attractors of two types.
\ The \emph{mixture }or \emph{spin glass states} have partial overlaps with
several patterns and thus lie close to the subspace spanned by the memory
patterns. \ \ \emph{Uncondensed states}, which have no counterpart in the
HN, are states in which none of the fields acting on the nodes are strong
enough to cause sign flips and the dynamics is therefore dominated by the
local potential. \ They have energies per node close to $-0.25$. \ The spin
glass states are intermediate in energy between the memory states and the
uncondensed states. \ \ In the range $\frac{1}{3}\leq\gamma\lessapprox1$,
pattern recognition by the network was found to be highly selective; \ the
input must be close to the stored pattern in order for the pattern to be
fully restored. \ The uncondensed states are numerous (of order $2^{N}$) \
and fill most of the configuration space. \ The memorized patterns and their
basins of attraction occupy isolated valleys among these states, but these
valleys expand as $\gamma$ increases. \ \ For $\gamma\gtrapprox1$, on the
other hand, the behaviour resembles that of the HN: \ there are no
uncondensed states, and the memory states have large basins of attraction.

In this paper we turn to the case where $p/N$ is of order unity. \ We are
interested in the maximum storage capacity, or the maximum number of
patterns that can be successfully stored and retrieved, as well as changes
in the network's performance as this limit is approached. \ Earlier work
with small networks \cite{IWANN} suggested that at least under certain
conditions the BGN could store many more patterns than the HN while
possessing few spurious attractors. \ It is known that the Hopfield storage
limit of $p\approx0.14N$ memory patterns can be exceeded if a more
complicated learning algorithm is used \cite{Amit2}, \ but in the BGN case
improved capacity is achieved with the familiar Hebb rule. \ Since the
previous results \cite{Chinarov}\cite{IWANN} were obtained with networks
much too small to be of practical interest (e.g., $N=5$), \ we now examine
larger networks, mainly through numerical simulations. \ (At the end of the
paper we will return briefly to the small-network case.) \ We find that the
high-loading behaviour, like that at low loading, depends strongly on $%
\gamma.$ \ For $\gamma\gtrsim1$, a Hopfield-like first-order phase
transition results in the destabilization of all memory states at a critical
value of $\alpha$. \ For $\gamma=2$ this transition occurs at $%
\alpha_{c}\approx0.1$, compared to $\alpha_{c}\approx0.14$ for the HN. \ For 
$\gamma=0.5$, on the other hand, \ we find that it is possible for the
stored patterns to remain stable even at loading factors of $%
\alpha\approx0.3 $ and higher. \ Furthermore, there is no sudden blackout; \
instead, the performance degrades gradually as $\alpha$ increases. \ The
price of this high capacity is that the retrieval of the patterns from
corrupted versions may be imperfect. \ 

The remainder of the paper is organized as follows: \ In section \ref%
{crosstalk} we discuss in general terms the effects of crosstalk, or
interference between different stored patterns. \ It is crosstalk which is
responsible for limitations on storage capacity. \ \ We compare the effects
of crosstalk in the BGN and the HN. \ This discussion provides a framework
for interpreting our numerical results. \ \ \ In section \ref{br} we examine
numerically the stability of memory patterns as a function of $\alpha$. \ We
find evidence of a first-order memory blackout phase transition in the BGN
at high $\gamma$, but not at low $\gamma$. \ In section \ref{basins}, \ we
examine the effects of high loading on the basins of attraction for the
memory states and on their retrieval from corrupted input. \ We see that
increasing the loading markedly alters the energy landscape. \ \ In section %
\ref{smallnet} we comment briefly on the behaviour of smaller networks and
on the relation between the previous small network results \cite{Chinarov}%
\cite{IWANN} and those we have obtained for larger networks. \ \ We conclude
with some discussion of the BGN results and possible directions for further
study. \ We make some conjectures concerning the performance of the BGN in
the presence of stochastic noise. \ 

\section{Pattern retrieval and crosstalk\label{crosstalk}}

In a given state $\mathbf{x}$, \ the input to the $i$-th node of the network
can be expressed as 
\begin{equation}
h_{i}=\gamma\sum_{j=1}^{N}w_{ij}x_{j}=\gamma\sum_{\mu=1}^{p}\xi_{i}^{\mu
}m_{\mu}-\gamma\frac{p}{N}x_{i},  \label{mField}
\end{equation}
using the Hebb rule (\ref{Hebb}) and the definition (\ref{mdef}) of $m_{\mu}$%
. \ In \cite{LowLoading}, we showed that when $\alpha\ll1$ there are stable
retrieval states in which\ $b_{\nu}=1$ for one particular $\nu$. \ For a
state in which the $\nu$-th overlap is large, we can decompose the field as
follows:%
\begin{align}
h_{i} & =\gamma\left( \xi_{i}^{\nu}m_{\nu}+\sum\limits_{\mu\neq\nu}\xi
_{i}^{\mu}m^{\mu}-\frac{p}{N}x_{i}\right)  \label{field} \\
& =\gamma\left( S_{i}+C_{i}-\frac{p}{N}x_{i}\right) .  \notag
\end{align}
We refer to the first term, $S_{i}$, as the \textquotedblleft
signal\textquotedblright\ term and the second term, $C_{i}$, as the
\textquotedblleft crosstalk\textquotedblright\ term. \ In the limit $\frac {p%
}{N}\rightarrow0$, \ the mutual overlaps between different patterns is
small, \ so that $m_{\nu}\ll1\;(\nu\neq\mu)$, and the last two terms in (\ref%
{field}) can be neglected. \ The signal term is then dominant and there is a
stable retrieval state given by $\mathbf{x}=\sqrt{1+\gamma}\mathbf{\xi }%
^{\nu}$ with $m_{\nu}=\sqrt{1+\gamma}$. \ We expect this solution to be
approximately valid for small but nonzero values of $\frac{p}{N}$. \ For
this case, \ the overlaps $m_{\mu}\;(\mu\neq\nu)$ behave as Gaussian
distributed random variables with zero mean and variance $1/\sqrt{N}$. \
Accordingly, the crosstalk term $C_{i}\equiv\sum\limits_{\mu\neq\nu}%
\xi_{i}^{\mu}m^{\mu}$ in (\ref{field}), being a sum of $p$ such quantities,
is a random quantity with zero mean and variance $\sqrt{p/N}$. \ \ The third
term, which arises from the subtraction of the diagonal elements, is of
order $p/N$ and thus is generally smaller than the crosstalk term. We will
neglect it for the moment.

In the absence of crosstalk, \ a retrieval state is not only linearly stable
(i.e., stable against small perturbations) but, for $\gamma>\frac{1}{3}$, it
is also stable against individual sign flips. \ \ The latter means that if a
retrieval state is corrupted by changing the sign of one or a small number $%
\ll N$ of nodes, then the dynamics will reverse the flipped sign and restore
the pattern. \ This happens because, in the absence of crosstalk, each node
experiences a field $\gamma S_{i}=\gamma\xi_{i}^{\nu}m_{\nu}$ which has the
same sign as $\xi_{i}^{\nu}$ and for $\gamma>\frac{1}{3}$ that field is
strong enough to overcome the potential barrier of the individual node. \ \
Now consider a given node (say, the $i$-th node) in the presence of
crosstalk. \ The crosstalk field acting on the $i$-th node may be either
aligned with or opposed to $x_{i}$. \ If it is aligned, then its effect on
that node is to increase the magnitude of $x_{i}$, making it larger than $%
\sqrt{1+\gamma}$. \ If it is opposed to $x_{i}$, then its effect is to \emph{%
decrease} the equilibrium magnitude of $x_{i}$. \ If the crosstalk term is
large enough, then it may be sufficient to overcome the local potential
barrier and reverse the sign of $x_{i}$, thus introducing a sign error into
the pattern. \ This will occur only if 
\begin{equation*}
C_{i}+S_{i}<-\frac{2\sqrt{3}}{9\gamma}.
\end{equation*}
By contrast, in the Hopfield model a sign error is introduced if 
\begin{equation*}
C_{i}+S_{i}<0.
\end{equation*}
Thus the relative strength of crosstalk required to introduce sign errors is
greater for the BGN than for the HN, and the discrepancy is greatest for
small values of $\gamma$. \ One might expect that this would make the BGN
less vulnerable to crosstalk (and the memory states more stable) \ than the
HN, especially at low $\gamma$, but this is not a foregone conclusion as the
BGN's dynamics include mechanisms which tend to amplify small initial
overlaps\cite{LowLoading} and could conceivably also amplify crosstalk. \
Our numerical results confirm that the BGN is in fact less prone to
crosstalk-induced errors at low values of $\gamma$, but not at high values.

\ \ If 
\begin{equation*}
-\frac{2\sqrt{3}}{9\gamma}<C_{i}+S_{i}<\frac{2\sqrt{3}}{9\gamma}
\end{equation*}
the crosstalk will not be strong enough to reverse the sign of $x_{i}$ if
initially $x_{i}$ is correctly aligned with $\xi_{i}^{\nu},$ but it will
nonetheless destroy the stability against a sign flip. \ In other words, if $%
x_{i}$ is initially misaligned, it will not be corrected. \ We may say that
the node is \emph{\textquotedblleft bistabilized\textquotedblright\ \ }but
not destabilized. \ (Such an effect cannot occur in the HN where the nodes
are not individually bistable.) \ Since the crosstalk is random, \ it will
in general bistabilize some nodes and not others, with the result that the
memory state will be stable against sign flips of certain nodes but not of
others. \ Thus, even though a memory state may be linearly stable, the
dynamics may only be able to correct some sign errors, not all. \ This
contrasts with the low-loading case where crosstalk is negligible and there
is a single threshold coupling strength, $\gamma=\frac{1}{3},$ above which
any single sign error can be corrected. \ In general, crosstalk results in
non-uniform behaviour among nodes, including different magnitudes of $x_{i}$
for different $i$.

\section{Stability of the memory states and remanent overlap\label{br}}

\ To examine the stability of the memory states,\ we followed a procedure
similar to that of reference \cite{Amit2}. \ Using an ensemble of
realizations of the random patterns, we made a number of trials in which the
network was initialized to the state $\mathbf{x}=\mathbf{\xi}^{\nu}$ for
some pattern $\nu$. \ The initial bit overlap $b_{\nu}$ was thus equal to 1.
\ We then allowed the state to evolve until an attractor was reached. \ In
each case we measured the final energy as well as the bit overlap between
the initial and final states, to which we refer as the \emph{remanent} bit
overlap $b_{r}$. \ We then constructed histograms showing the probabilities $%
P(b_{r})$ for $b_{r}$ falling in intervals of width 0.05. \ Each of the
histograms discussed in this section represents an ensemble of at least 200
initial conditions. \ \ If the memory state is entirely stable, as is the
case at very low loading, then after the dynamics converges the final bit
overlap will still be equal to $1$. \ \ For higher loading, however, \ the
crosstalk fields may introduce sign errors so that $b_{r}<1$. \ 

\subsection{HN and BGN with high $\protect\gamma$}

\begin{figure}
[ptb]
\begin{center}
\includegraphics[
height=3.691in,
width=4.6951in
]%
{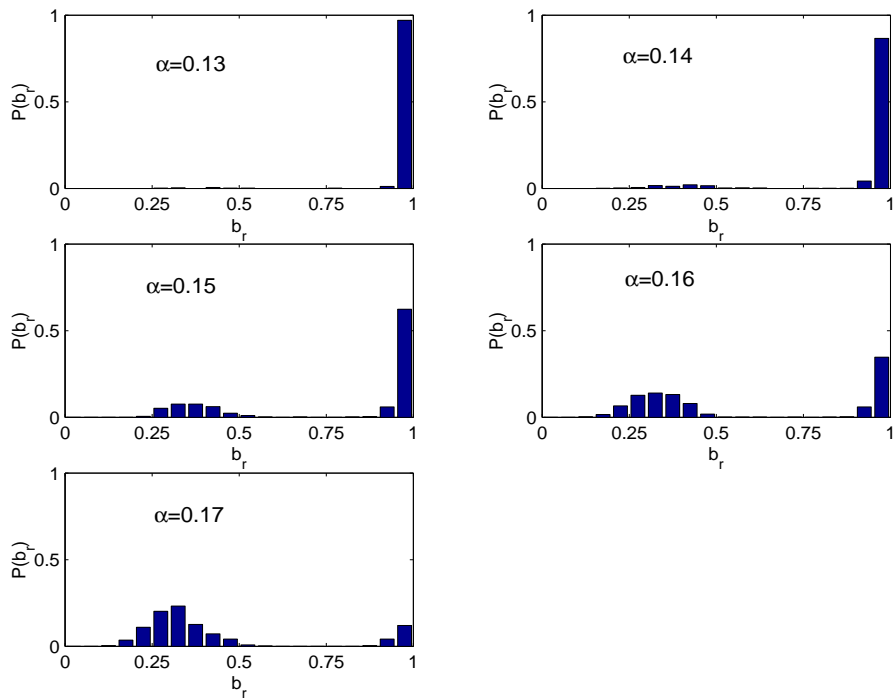}%
\caption{A series of
remanent overlap histograms for a HN with N=2000 nodes, at different values
of the loading fraction $\protect\alpha \equiv p/N$. \ \ When the critical
loading $\protect\alpha _{c}\protect\cong 0.148$ is exceeded, the strong
peak at $b_{r}>0.95$ decays and another peak appears at. $b_{r}\approx 0.3$}%
\label{ohisth2000}%
\end{center}
\end{figure}

\begin{figure}
[ptb]
\begin{center}
\includegraphics[
height=3.6919in,
width=4.6951in
]%
{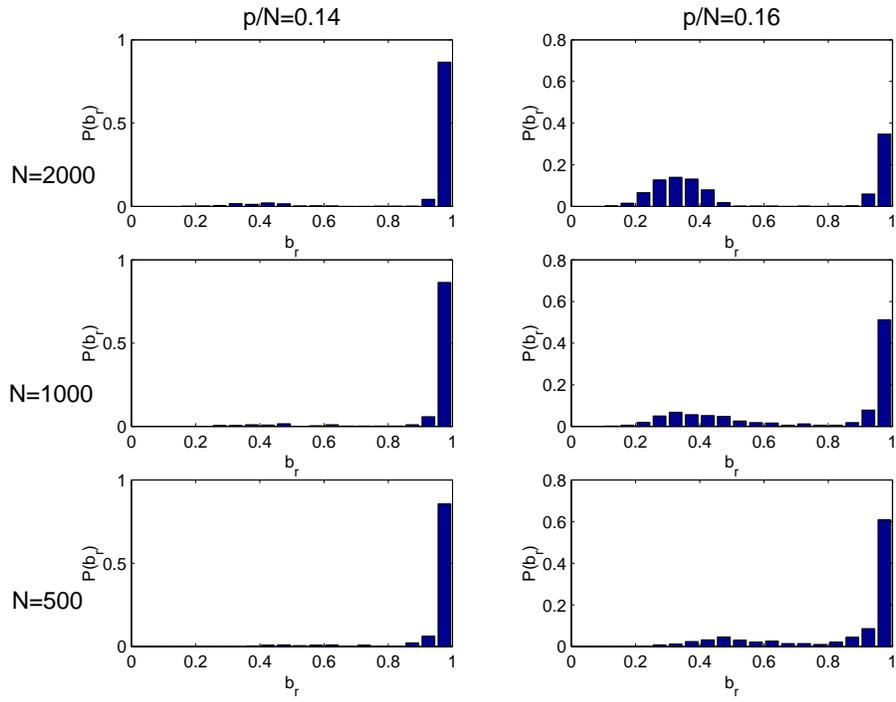}%
\caption{Remanent overlap
histograms for the Hopfield network below the critical loading (left column)
and above (right column), showing finite size effects. \ The transition
becomes sharper as $N$ increases. \ }%
\label{ohisth}%
\end{center}
\end{figure}

First consider the case of the HN$.$ \ Figure \ref{ohisth2000} shows a
series of $P(b_{r})$ histograms for the HN at different values of $\alpha
=p/N$. \ (The data here are our own, but the results are comparable to those
given in \cite{Amit2}. We include them for comparison with BGN results.) \
At lower levels of loading, crosstalk introduces few errors, so $b_{r}>0.95$
in the large majority of cases. \ \ However, as $\alpha $ increases beyond
the critical value $\alpha _{c}\approx 0.148$, \ the high-$b_{r}$ \ peak of\
the distribution begins to vanish and a second, broader peak begins to grow
in the vicinity of $b_{r}\approx 0.3$. \ The states in the second peak are
spin glass states. \ As shown in figure \ref{ohisth}, this transition
becomes sharper with increasing network size, and finite-size scaling
analysis shows behaviour characteristic of a first order phase transition in
the thermodynamic limit \cite{Amit2}. In the limit $N\rightarrow \infty $
the associative memory fails suddenly as the critical loading is exceeded---
the remanent overlap drops abruptly from near 1 to 0.3. \ This nonzero value
of the remanent overlap above the critical loading was noted in \cite{Amit2}
and attributed to replica symmetry breaking, as the replica symmetric theory
predicts that $b_{r}$ should drop to zero above the phase transition. \ This
phenomenon is related to the non-zero remanent magnetization of a spin glass
\ \cite{Mezard}.
\begin{figure}
[ptb]
\begin{center}
\includegraphics[
height=3.6616in,
width=4.6951in
]%
{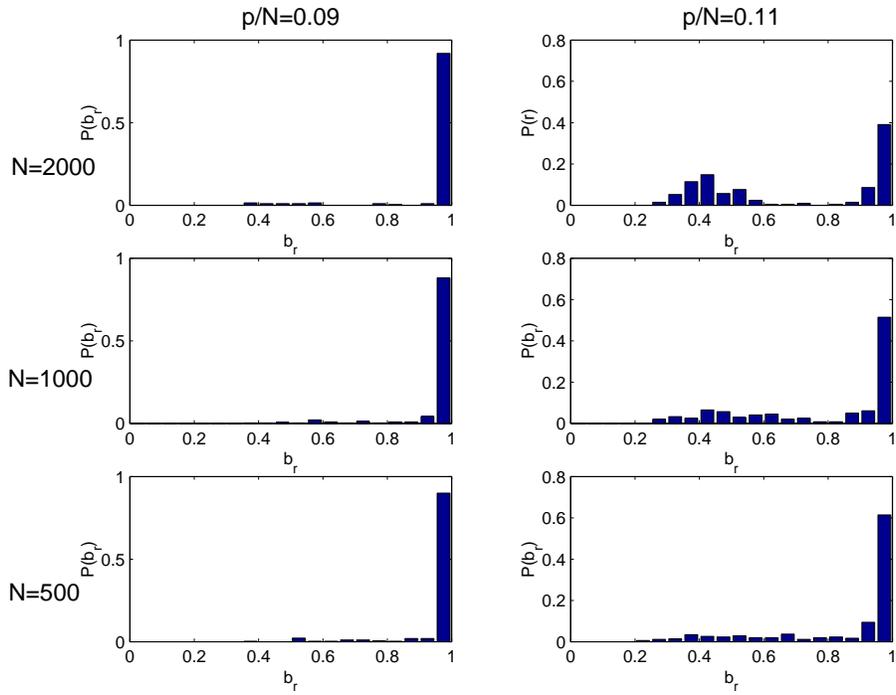}%
\caption{Remanent overlap histograms for the
BGN with $\protect\gamma =2$ show evidence of a first-order transition at $%
0.9<\protect\alpha _{c}<0.11$. \ \ Note the similarity with figure \protect
\ref{ohisth}.}%
\label{ohist2}%
\end{center}
\end{figure}

In the BGN with $\gamma=2$, \ a similar transition evidently occurs at $%
\alpha_{c}\approx0.1$. \ As evidence, in figure \ref{ohist2}\ we show two
series of histograms at increasing $N$, one below the suspected transition
and one above. \ As in the HN case, the transition grows sharper with
increasing network size. \ Below the critical loading, the high-$b$ peak
remains robust as $N$ increases, \ while above the critical loading the high-%
$b$ peak shrinks with increasing network size and the low-$b$ peak grows.
Two quantitative differences are that the critical loading $\alpha_{c}$ is
lower in the BGN case, $\alpha_{c}\approx0.1$, \ while the average remanent
overlap above the critical loading is higher, near 0.45.

\begin{figure}
[ptb]
\begin{center}
\includegraphics[
height=4.5558in,
width=4.305in
]%
{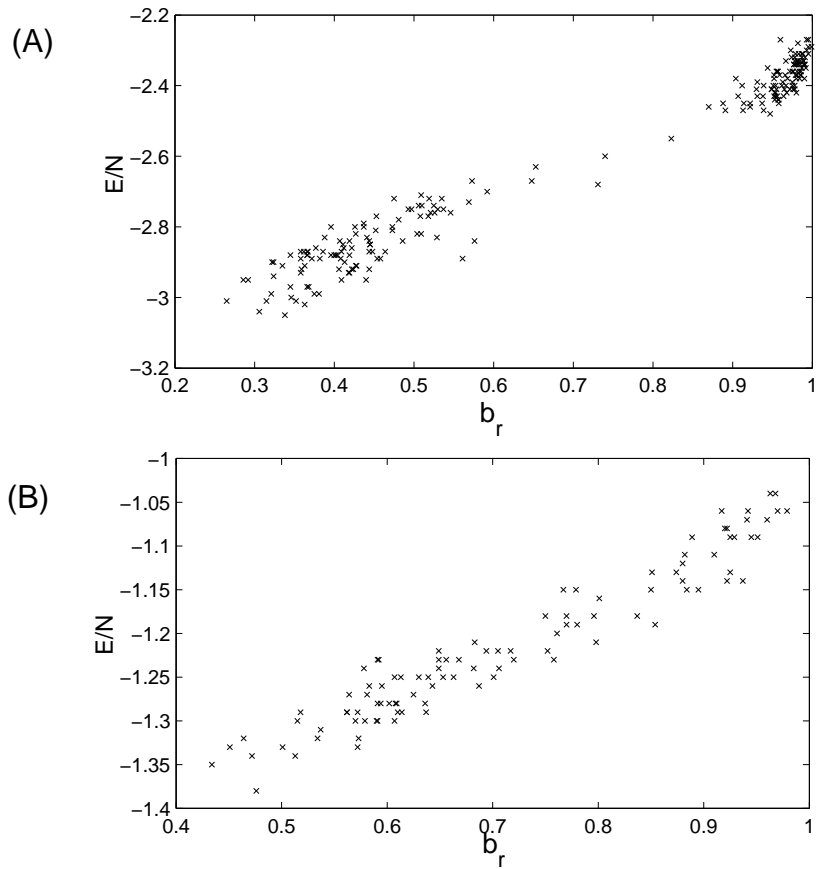}%
\caption{Scatter plots of final energies and remanent
overlaps for trajectories starting from memory patterns, for BGN above
critical loading. \ (A) BGN with $N=2000$, $\protect\gamma=2$ and $\protect%
\alpha=p/N=0.11$. \ Note that the overlap is strongly correlated with the
energy and there is a gap in energy between the high and low-$b_{r}$ states.
\ This gap represents the latent heat of the phase transition. \ \ (B) \ BGN
with $N=2000$, $\protect\gamma=1$ and $\protect\alpha=0.2$. \ As in the $%
\protect\gamma=2$ case, the energy is strongly correlated with $b_{r}$, but
the high and low groups overlap and the energy difference is smaller.}%
\label{scatcombined}%
\end{center}
\end{figure}

The first-order nature of the transition is confirmed by examining the
energies of the final states. \ These energies and the overlaps are shown in
a scatter plot in figure\ \ref{scatcombined}A. \ The spin glass states
associated with the low-$b_{r}$ peak are clustered at energies below those
of the retrieval states. \ The gap in energy between these two clusters
corresponds to the latent heat of the phase transition. \ 
\subsection{BGN at low $\protect\gamma$}

\begin{figure}
[ptb]
\begin{center}
\includegraphics[
height=3.8078in,
width=4.6951in
]%
{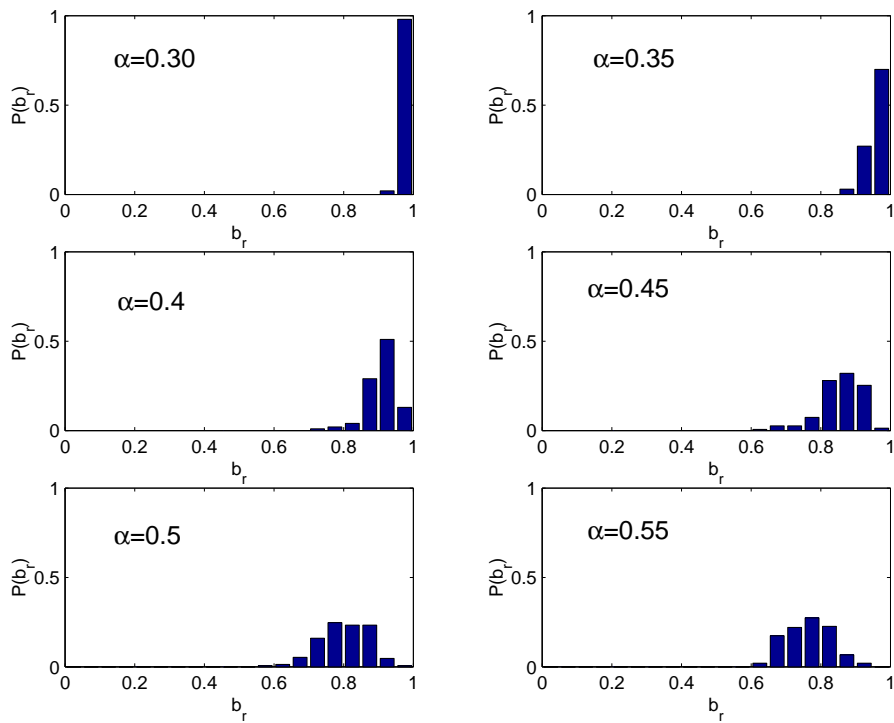}%
\caption{Remanent bit
overlap histograms for the BGN with $N=1000$ and $\protect\gamma=0.5$, \ at
a series of increasing values of the loading factor $\protect\alpha$. \ Few
errors occur even at $\protect\alpha=0.3$, \ and the memory degrades
gradually rather than abruptly with increasing $\protect\alpha.$}%
\label{ohist05}%
\end{center}
\end{figure}

For $\gamma=0.5$, in contrast to $\gamma=2$, the BGN's behaviour differs
markedly from that of the HN. \ A series of $P(b_{r})$ histograms for
different values of $p/N$ is shown in figure \ref{ohist05}. \ Two features
are evident. \ First, the stored patterns remain stable with few errors even
up to high levels of storage, \ $\alpha=0.3$. \ Second, there is no evidence
of a discontinuous memory failure; \ rather, the retrieval quality as
measured by $b_{r}$ appears to degrade continuously as $\alpha$ increases. \
No second peak appears in the histograms; instead, the high-$b$ peak first
spreads and then begins to drift downward as errors accumulate. \ 

At an intermediate value $\gamma=1$ the $P(b_{r})$ histograms \ (figure \ref%
{ohist1}) suggest a first-order transition, although the evidence is less
pronounced than in the $\gamma=2$ case. A second peak appears above the
transition, and the high-$b_{r}$ peak shows clear signs of shrinking as $N$
increases, but the tails of the two peaks overlap substantially . \ The
greater overlap between the peaks comes about for two reasons. \ First, the
high-$b_{r}$ peak above the critical loading is broader than in the $%
\gamma=2 $ case. \ Second, the remanent magnetisation is much higher (i.e.,
the drop in $b_{r}$ at the critical point is much smaller.) \ The latent
heat is also much smaller, as can be seen from the scatter plot of the
energy (figure \ref{scatcombined}B). The critical loading, or storage
capacity, is approximately 0.17, higher than for $\gamma=2$ and higher than
for the HN. \ \ 
\begin{figure}
[ptb]
\begin{center}
\includegraphics[
height=4.5455in,
width=6.0502in
]%
{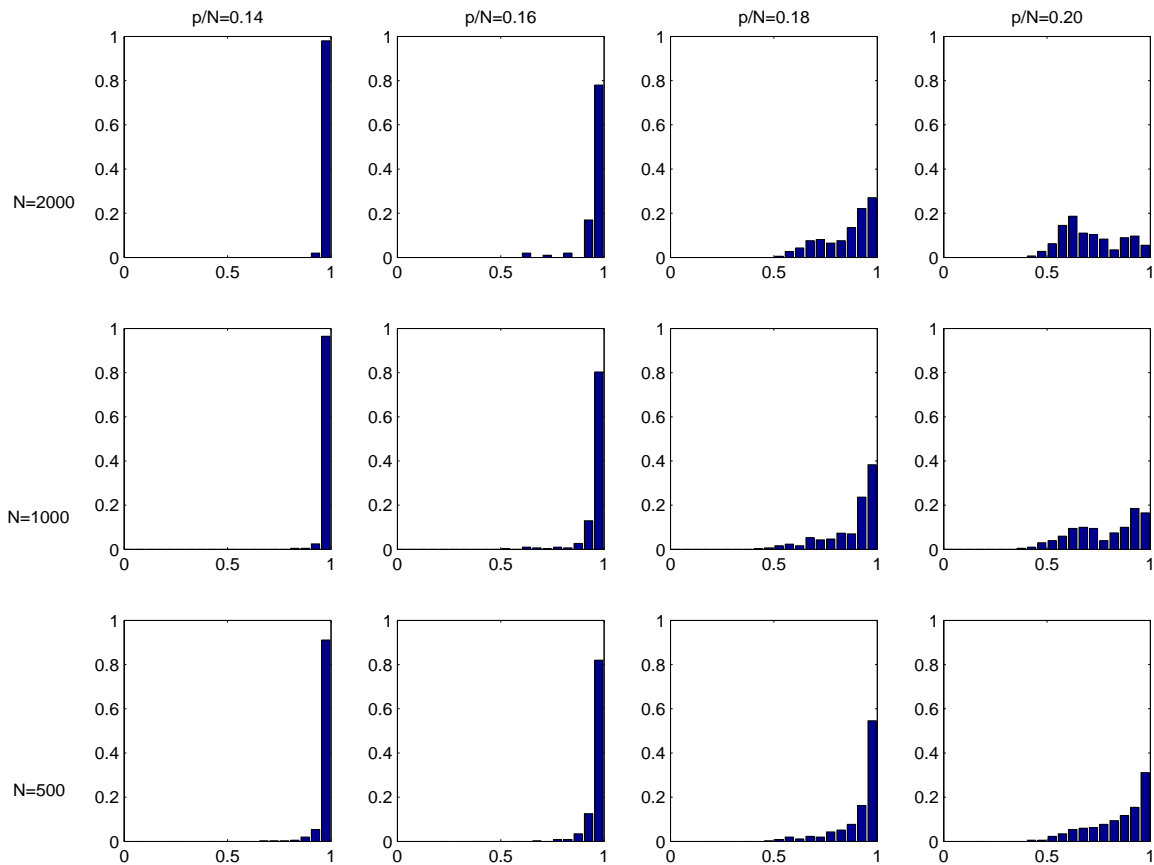}%
\caption{Remanent overlap histograms for BGN with $%
\protect\gamma=1$ are consistent with a first-order transition at $0.16<%
\protect\alpha_{c}<0.18.$ \ }%
\label{ohist1}%
\end{center}
\end{figure}
\section{Attractors, basins and the energy landscape at high loading\label%
{basins}}

In the previous section, we examined the trajectories of initial conditions
corresponding to memorized patterns and determined whether these
trajectories remain close to the pattern or move away from it. \ Such
experiments, however, probe only one aspect of network performance. \ We are
interested not only in the stability of memory states but also in the sizes
of their basins of attraction. \ The function of associative memory depends
on the ability of the dynamics to correct partially corrupted patterns. \
More generally, we are interested in the evolution of the energy landscape
with increasing $\alpha$. \ To address these issues, we performed two
additional sets of numerical experiments. First, we examined the attractors
reached from a large number of \emph{random} initial conditions to obtain a
uniform sampling of the phase space and a broad picture of the energy
landscape. \ Second, we examined the fate of initial conditions at specified
Hamming distances from memory patterns. \ The latter set of experiments
probes the landscape in the vicinity of the memory states. \ Results of
similar experiments were given in ref. \cite{LowLoading} for the case of low
loading. \ 

\subsection{Evolution of the energy landscape: attractors reached from
random initial states}

\begin{figure}
[ptb]
\begin{center}
\includegraphics[
height=3.352in,
width=4.1433in
]%
{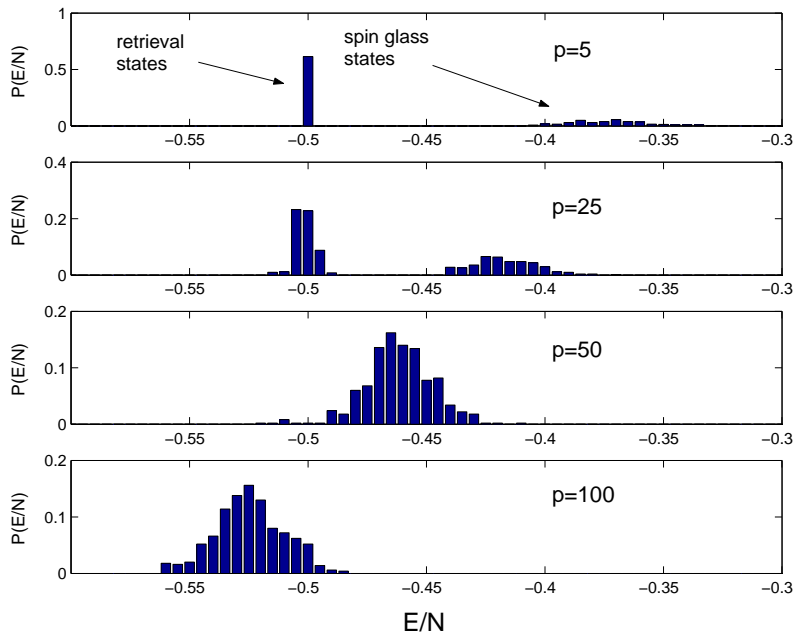}%
\caption{Energy histograms of $P(E/N)$ for attractors reached from random initial
states of the HN with $N=1000$ nodes. \ As the number $p$ of stored patterns
increases, \ the peak at $E/N=-0.5$ corresponding to the retrieval states
shrinks (and also spreads slightly) while the the spin glass states drift
downward in energy until they are lower than the retrieval state energies. \  }%
\label{seedH}%
\end{center}
\end{figure}

\begin{figure}
[ptb]
\begin{center}
\includegraphics[
height=3.3347in,
width=4.1433in
]%
{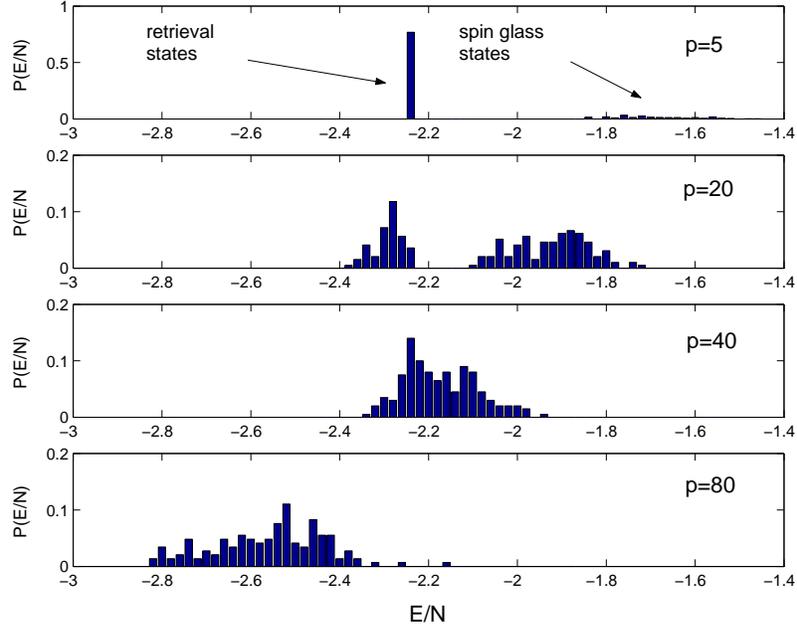}%
\caption{Attractor energies from random initial
conditions for BGN with $\protect\gamma=2.$ \ Qualitative behaviour
resembles that of the HN as shown in figure \protect\ref{seedH}. \ One
difference is that the spreading of the retrieval state energies with
increasing loading is more pronounced. \ }%
\label{seed2}%
\end{center}
\end{figure}

\begin{figure}
[ptb]
\begin{center}
\includegraphics[
height=3.3529in,
width=4.1433in
]%
{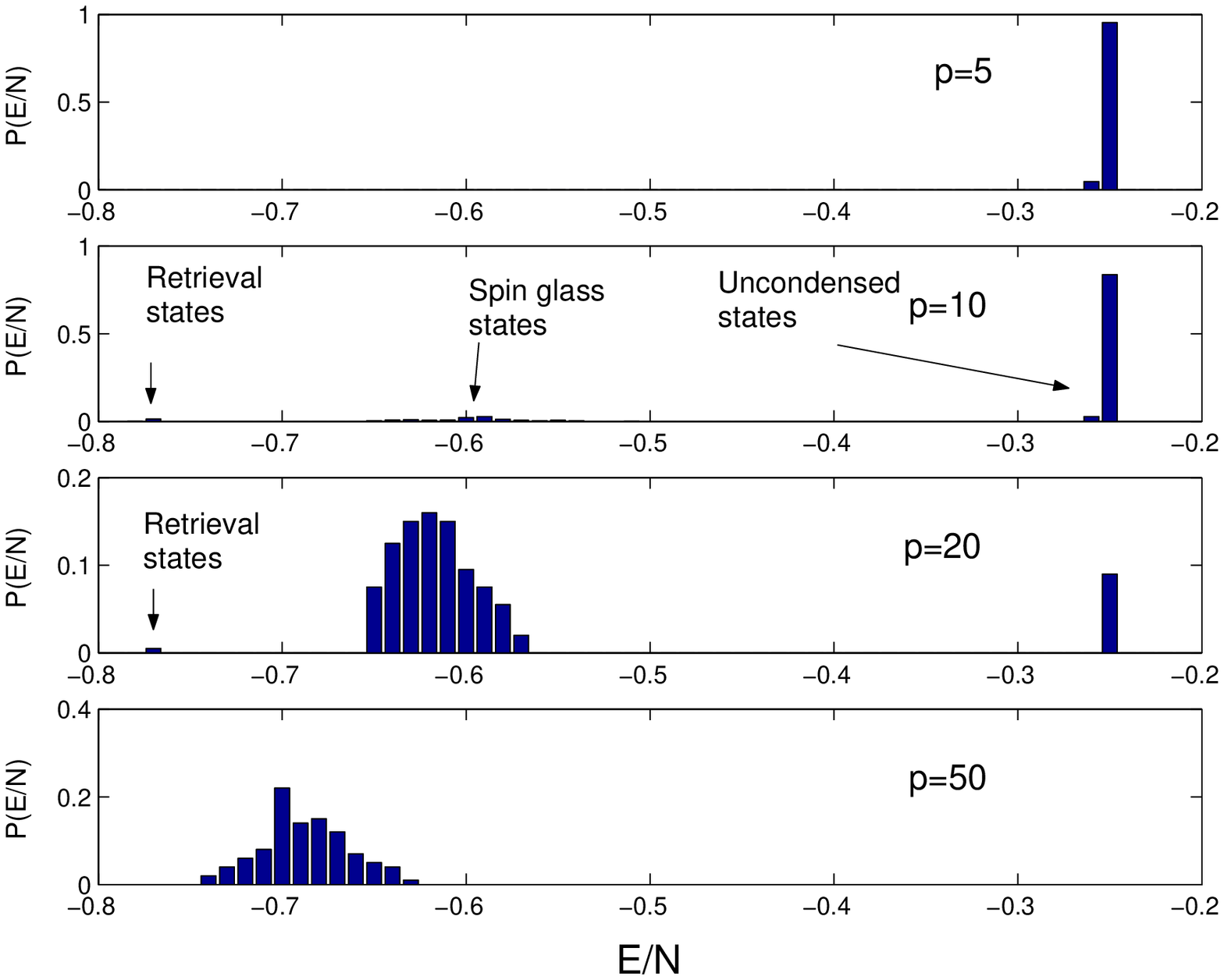}%
\caption{Attractor energies from random initial conditions for BGN with $N=1000$
and $\protect\gamma=0.75.$ For low loading, the uncondensed states (with
energy $E/N\approx-0.25 $) \ fill most of the configuration space. \ The
other two peaks in the histogram are very small. \ As the loading increases,
however, the uncondensed states disappear. \ As in all cases, the
\textquotedblleft spin glass\textquotedblright\ peak grows larger with
increasing $\protect\alpha$ and shifts downward in energy.}%
\label{seed075}%
\end{center}
\end{figure}

In the case of the HN, \ it is known that in the thermodynamic limit the
memory states are the lowest energy states for $\alpha<0.05$, while for
higher values of $\alpha$ spin glass states arise which have lower energies.
\ Up to $\alpha_{c}\approx0.148$ the memory states remain local minima of
the energy even though they are not the global minima. \ Above $\alpha_{c}$
they cease even to be local minima and therefore become unstable\cite{Amit2}.%
\footnote{%
For a schematic illustration of the evolution of the energy landscape, \ see
fig. 2.18 of reference \cite{HKP}. \ \ } \ The drop in energy from the
memory states to the spin glass states at $\alpha_{c}$ is the latent heat. \
\ One way to observe the evolution of the energy landscape is to examine the
attractors reached from an ensemble of random initial conditions which
effectively samples the configuration space. \ Figures \ref{seedH}-\ref%
{seed075} show histograms for the energies of attractors sampled in this
manner. \ In each case, we sampled a total of at least 200 random initial
conditions with several realizations of the random patterns\textbf{\ }$%
\mathbf{\xi}^{\mu}$. \ \ In figure \ref{seedH}, for the HN, we can see that
at low loading the attractors are separated into two groups, the retrieval
states at $E/N=-0.5$ and spurious states at a range of higher energies. \
The probability of retrieving a memory state from a random initial condition
is high. \ With increasing $\alpha$, \ the spurious states move to lower in
energies until they are below the memory states. \ At the same time, their
basins of attraction take up a larger portion of the configuration space, \
as is apparent from the growing size of the spin glass peak in the histogram
and the shrinking size of the retrieval state peak. \ \ Figure \ref{seed2}
shows that for the BGN with $\gamma=2$ the evolution is qualitatively
similar. \ In figure \ref{seed075}, for $\gamma=0.75,$ we observe that an
additional effect of high loading is to destabilize the uncondensed states.
\ \ At low loading, the uncondensed states dominate the configuration
space---almost all random initial conditions land on an uncondensed state,
as was noted in \cite{LowLoading}. \ The other two types of state begin to
show their presence as the loading is increased, while the uncondensed
states eventually disappear. \ \ 

\subsection{Retrieval of patterns from corrupted versions}

To obtain information about the landscape in the vicinity of the memory
states and about the shapes of their basins of attraction, we examined the
fates of initial conditions which were not random but rather at specified
initial overlaps with particular stored patterns. \ As in \cite{LowLoading},
these initial configurations were generated by starting with a particular
\textquotedblleft target\textquotedblright\ pattern and flipping the signs
of a specified number of randomly chosen nodes. \ For each value of the
initial overlap $b_{init}$, we generated an ensemble of initial conditions
for several different realizations of the random set of memory patterns. We
then evolved these states until the dynamics converged, and evaluated $%
b_{final}$, the final overlap with the target pattern, for each trajectory.
\ If $b_{final}=1$, this signifies that all signs which were initially
flipped have been corrected and the target pattern has been retrieved
perfectly. \ If $b_{init}<b_{final}<1$, \ then the pattern has been
retrieved imperfectly. \ The final state is closer to the stored pattern,
but not all errors have been corrected. \ \ If $b_{init}>b_{final}$, then
the trajectory has moved farther away from the stored pattern. \ As
discussed in section \ref{crosstalk}, the ability of the network to correct
sign errors depends on the competition among the signal, the crosstalk, and
the local potential. \ The initial states currently discussed have at least
a moderately large overlap with the target pattern, resulting in a signal,
and random overlaps with the other memory patterns, resulting in crosstalk.
\ 

Consider now the HN. \ At low loading, there is very little crosstalk. \ The
crosstalk term is typically of order $\sqrt{p/N}$. \ If the HN is set in an
initial condition whose overlap with the target pattern $\mathbf{\xi}$ is $%
b_{init}>\sqrt{p/N}$, \ then most nodes experience a net field which is
aligned with the target pattern. \ Nodes which are initially misaligned with
the target state ($x_{i}=-\xi_{i}$) will tend to change their signs and
align with $\xi_{i}$. \ Each node which realigns in this way increases the
value of $b$ and thus increases the strength of the signal acting on the
remaining nodes. \ \ As a result, even if in the initial state some fraction
of the nodes experience a net field opposed to $\xi_{i}$, eventually the
growing signal may overcome the crosstalk and correct those nodes as well. \
Therefore for low loading, as long as the initial state has $b_{init}>\sqrt{%
p/N}$, the probably of completely retrieving the target pattern is close to
unity. \ As the loading ratio $\alpha$ increases, however, \ the typical
crosstalk becomes stronger, and a higher signal is required to overcome the
crosstalk noise. \ Therefore sign errors are not likely to be corrected
unless the initial overlap is above a threshold, which grows higher with
increasing $\alpha$. \ If the crosstalk is too large, then some signs which
are initially aligned with the pattern may be flipped, and the state may
move away from the target pattern instead of toward it. \ Each node which
flips out of alignment with the target pattern reduces the size of the
overlap and hence of the signal, which makes other nodes more susceptible to
crosstalk-induced errors, and a cascade of errors can occur. \ The critical
loading $\alpha_{c}$ is the level at which even a state with $b_{init}=1$
becomes unstable against such a cascade of errors. \ 

\begin{figure}
[ptb]
\begin{center}
\includegraphics[
height=4.8317in,
width=4.6613in
]%
{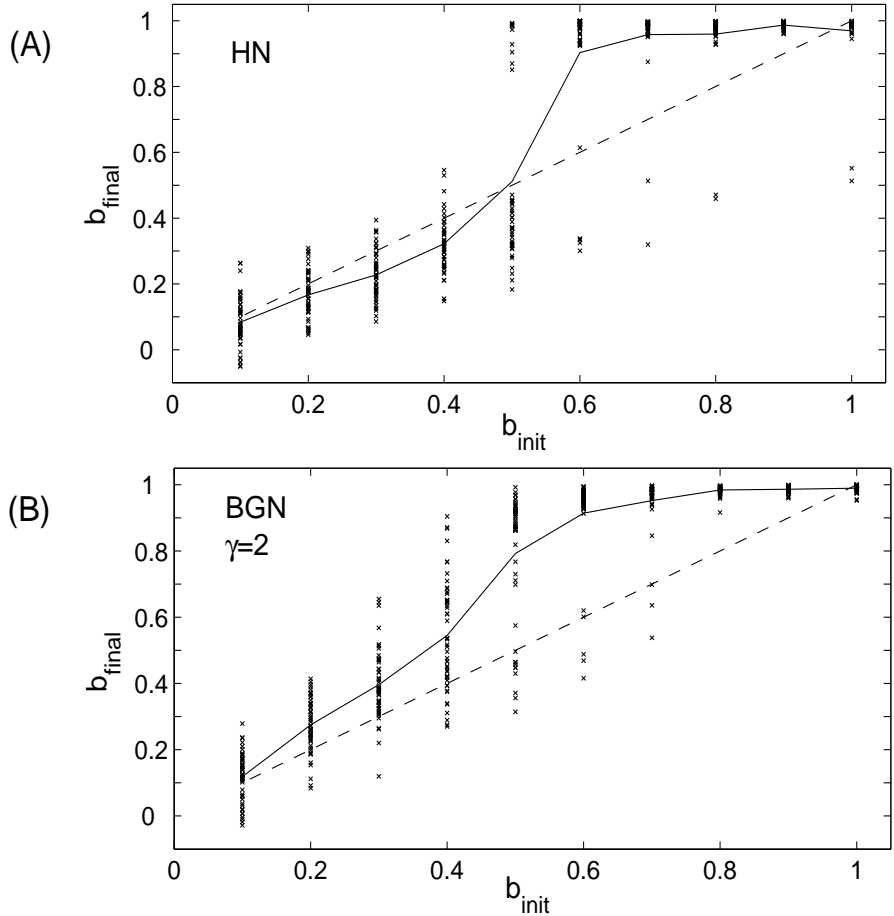}%
\caption{Scatter plots of $b_{final}$
vs. $b_{init}$ for networks slightly below critical loading. \ Points show $%
b_{final}$ for an ensemble of initial conditions with specified $b_{init}$.
\ \ The average of $b_{final}$ is shown by the solid curve. The dotted
diagonal line $b_{final}=b_{init}$ is drawn for comparison: \ points above
the line have $b_{final}>b_{init}$.\ \ (A) HN\ with $N=2000$, $p=260$. \ (B)
\ BGN with $N=2000$, $\protect\gamma=2$ and $p=170$. \ In both cases, the
retrieval quality as measured by $b_{final}$ drops sharply for $b_{init}<0.6$%
. \ For the BGN, however, the average $b_{final}$ is always greater than $%
b_{init}$. \  }%
\label{boundscatH2}%
\end{center}
\end{figure}

Figure \ref{boundscatH2}A shows a scatter plot of $b_{final}$ vs. $b_{init}$
for a HN slightly below critical loading. There is a threshold overlap for
retrieval. \ If $b_{init}>0.5$ the signal is strong enough to correct errors
and the majority of trajectories flow towards the target pattern. \ For $%
b_{init}<0.5$ the majority of trajectories move instead away from the target
pattern. \ As $\alpha$ increases, the threshold value of $b_{init}$ for
retrieval increases until at $\alpha_{c}$ it reaches 1. \ \ The plot in
figure \ref{boundscatH2}A is for a network with $N=2000$; \ experiments with
networks of different sizes reveal that the retrieval threshold becomes
sharper as $N$ increases. \ 

In the BGN, on the other hand, the dynamics of error correction is more
complicated due to the local potential. \ At strong coupling $\gamma\gtrsim1$%
, the potential barriers against sign flips are less important than at weak
coupling. \ As a result the BGN in this regime behaves in many respects like
the HN. \ It is not surprising, then, that the scatter plot of $b_{final}$
vs. $b_{init}$ for a BGN with $\gamma=2$ slightly below its critical loading
(figure \ref{boundscatH2}B) appears qualitatively similar to the
corresponding figure \ref{boundscatH2}A for the HN. \ There is a threshold
(approximately $b_{init}=0.6$) \ below which the probability of fully
retrieving the target pattern drops sharply. \ Above this threshold, the
signal is evidently strong enough to correct most sign errors. \ A key
difference, however, is that even below this threshold the average $%
b_{final} $ is larger than $b_{init}.$ \ This means that the majority of a
trajectories move toward the target pattern rather than away from it, \ but
only some of the sign errors are corrected, not all. \ 

\begin{figure}
[ptb]
\begin{center}
\includegraphics[
height=5.137in,
width=6.0528in
]%
{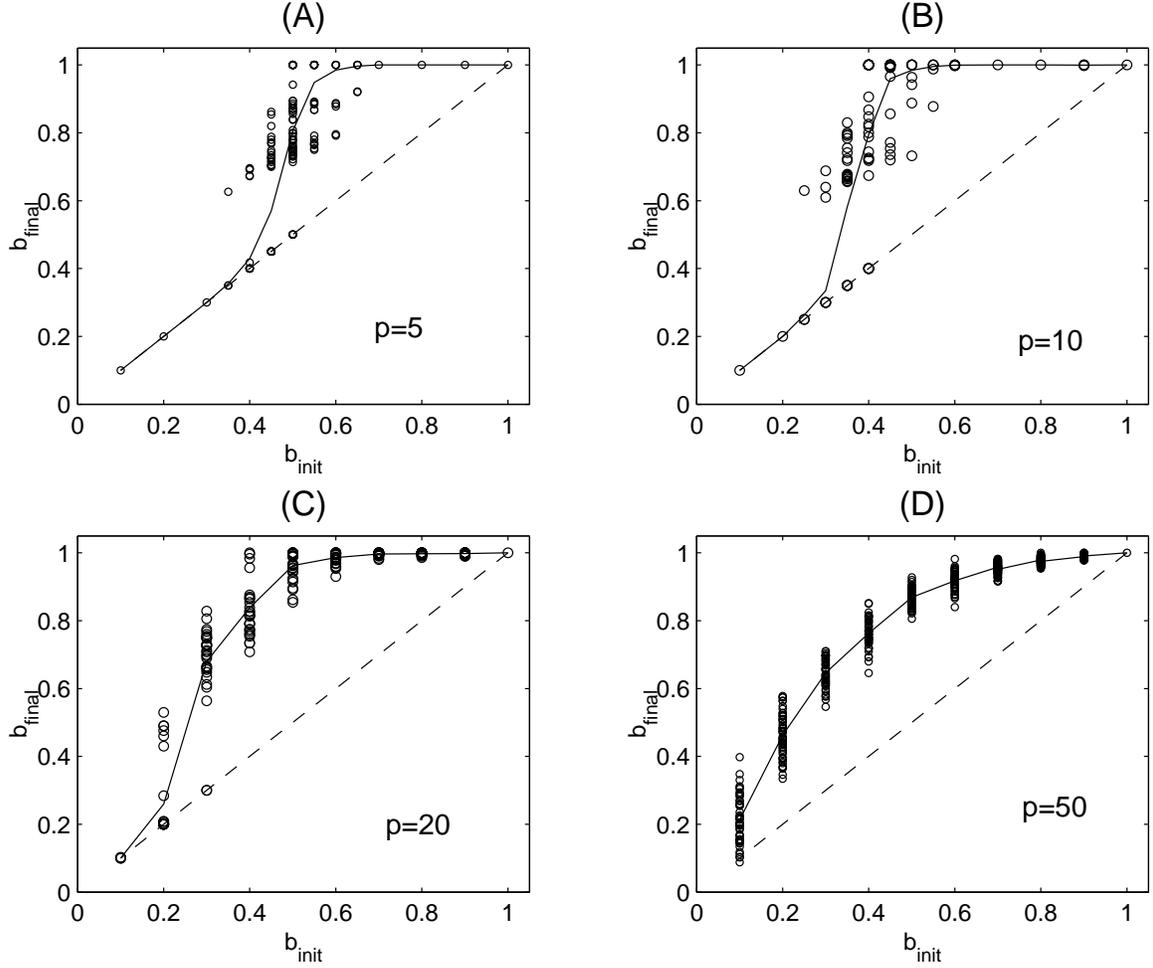}%
\caption{Scatter plots showing pattern
retrieval by the BGN with $N=1000$, $\protect\gamma =0.5$. \ As in figure 
\protect\ref{boundscatH2}, the solid curve shows the average value of $%
b_{final}$. \ Uncondensed states lie on the line $b_{final}=b_{init}$: \ no
sign flips occur and $b$ maintains exactly its initial value. \ (A) $p=5$ \
(B) $p=10$ \ (C) $p=20$ (D) $p=50.$ }%
\label{scat05}%
\end{center}
\end{figure}

In the case of the BGN with $\gamma=0.5$, the local potential barriers are
important, and interesting dynamics results from the competition among the
signal from the target pattern, the crosstalk from other patterns, and the
local potential. \ In the $\alpha\ll1$ case \cite{LowLoading}, sign errors
can be corrected only if the signal is strong enough to overcome the local
potential barriers, \ and therefore there is a threshold value of the
initial overlap which is much larger than $\sqrt{p/N}$. For the case $N=1000$%
, $p=5$, for example, we find that this threshold is approximately $%
b_{init}=0.5$. \ For $b_{init}>0.5$ there is a large probability that the
target pattern is retrieved perfectly, while for $b_{init}<0.5$ there is a
large probability that the network will be stuck in an uncondensed state
with $b_{final}=b_{init}$. \ In the intermediate range $0.4<b_{init}<0.6$,
there is also a significant probability that the trajectory is attracted to
an asymmetric spurious state in which $b_{final}$ is large but not unity and
there are larger than random overlaps with one or more other memory
patterns. \ \ This behaviour is illustrated by the scatter plot of figure %
\ref{scat05}A. \ As the loading increases \ (Figure \ref{scat05}A-D),
something surprising happens: \ at first, the basins of attraction of the
memory states \emph{expand} slightly, \ contrary to what one would expect
from the HN. \ \ The frontier of the uncondensed states is pushed back to
lower values of $b_{init}$. \ For $p=50$, or $\alpha=0.05$ (fig. \ref{scat05}%
D) \ almost all states with $b_{init}>0.1$ undergo some motion toward the
target pattern, even if the pattern is not retrieved perfectly. \ As $p$
increases further, the $b_{final}$ vs. $b_{init}$ plot preserves the
approximate shape of fig. \ref{scat05}D. \ 

\begin{figure}
[ptb]
\begin{center}
\includegraphics[
height=3.8735in,
width=5.1526in
]%
{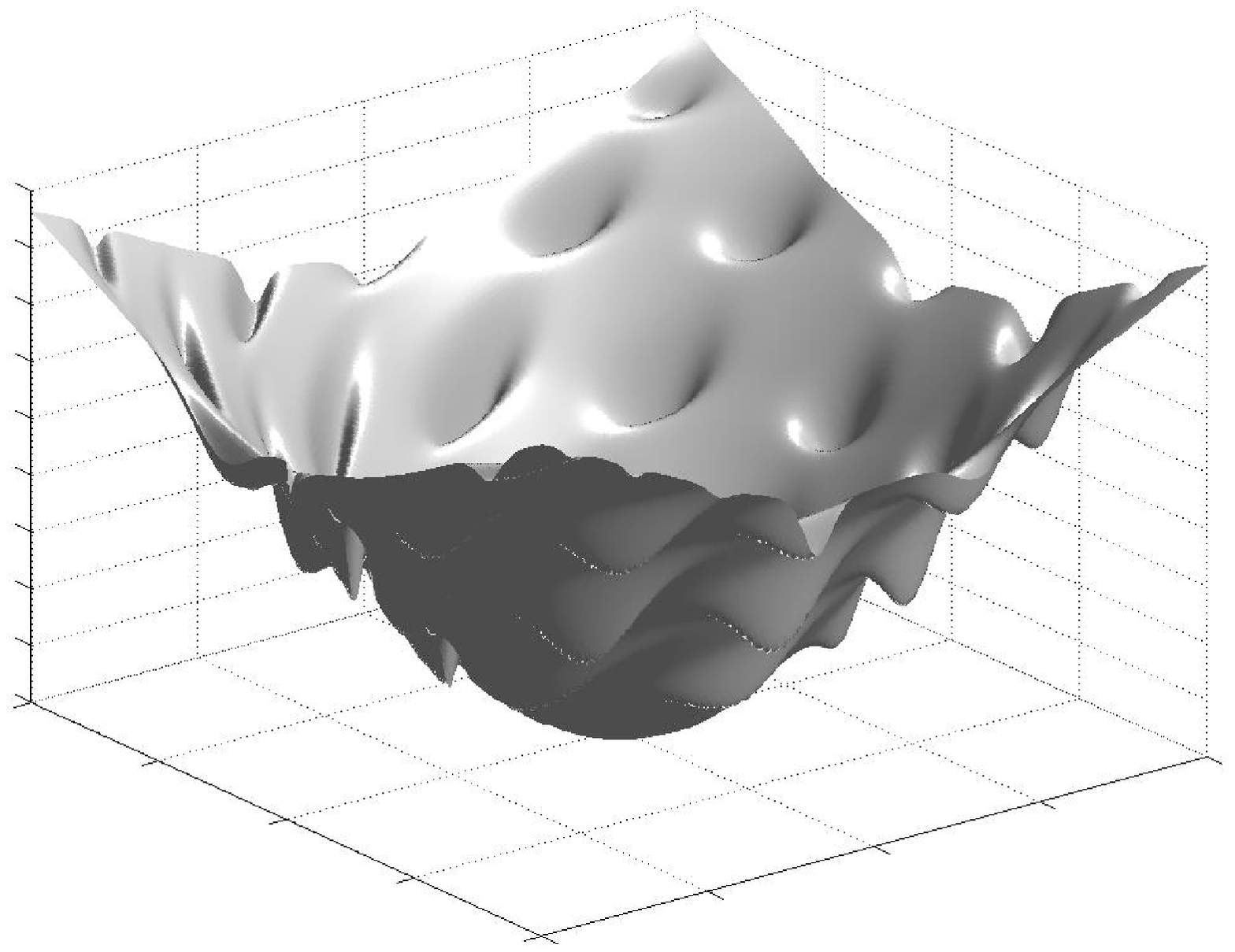}%
\caption{Schematic picture of
a funnel, or bumpy basin. \ Most trajectories travel some distance toward
the global minimum (representing for example a memory state) but become
trapped in a local minimum (a state with some errors) before reaching the
bottom. \  }%
\label{bumpy}%
\end{center}
\end{figure}

The dynamics of retrieval and error correction for the BGN with weak
coupling and high loading is evidently quite different from that of the HN.
\ In the $\alpha =0.05$ case of figure \ref{scat05}D, $b_{final}$ is always $%
1$ if $b_{int}=1$, which means that the memory state is stable. \ However,
it is retrieved only imperfectly when sign errors are introduced: if $%
b_{init}<1$ then \ $b_{init}<b_{final}<1$. \ This indicates that some nodes
remain bistable and cannot be corrected. \ If the initial state is close to
the target pattern, then the majority of errors are corrected, but that
fraction decreases with decreasing $b_{init}$. \ \ Such a partial correction
of errors does not often occur in the HN case. \ In the latter case, an
initial condition either flows all the way to the retrieval state or moves
away from it toward another attractor. \ The retrieval state may itself have
a small number of errors due to crosstalk, but the presence of these errors
does not depend on the initial state. \ The energy landscape of the HN in
the neighborhood of a memory state apparently has the shape of a smooth
basin --- once the basin is entered, \ the trajectory usually runs without
obstruction to the attractor at the bottom. \ For the weakly coupled BGN, on
the other hand, the landscape appears to have the structure of a
\textquotedblleft funnel,\textquotedblright\ \cite{funnel1}\cite{funnel2}%
\cite{Karplus} i.e., a sequence of local minima at decreasing energies, with
low potential barriers separating each state from the next. \ There is a
region of configuration space which has an overall tilt toward the retrieval
state, but in which there are many local minima which may trap the
trajectory before it reaches the retrieval state. \ This is illustrated
schematically in figure \ref{bumpy}. \ Funnel-shaped energy landscapes were
first suggested in the context of protein folding dynamics. \ \ \ \ \ \ \ \

\section{A Comment on Small Networks\label{smallnet}}

So far, this paper and the companion paper \cite{LowLoading} have focused
mainly on large networks of $N=1000$ or more. \ However, \ some applications
of neural network algorithms to robotics and other areas make use of
networks of only 20 or 100 nodes. \ Experimental studies of BGN-like
chemical reactor networks \cite{ChemNetE}\cite{ChemNetM} used fewer than 10
nodes. \ The current work on the BGN was inspired in part by results
suggesting that the BGN could store many more patterns than the HN, with
fewer spurious attractors \cite{IWANN}. \ The latter results were inferred
from a few selected cases using very small networks, \ and so we attempted
to test the genericness of these results for a variety of small networks as
well as for larger networks. \ 

It should be noted that with very small networks, there are great
fluctuations in properties depending on the particular set of memory
patterns chosen, \ as it is impossible to ignore the mutual correlations
among patterns. \ Results for the maximum storage capacity of a network are
well-defined only in the thermodynamic limit $N\rightarrow\infty$, and for
small $N$ even an HN may in particular cases be able to store more than $%
0.14N$ stable patterns. \ For these reasons one cannot draw strong general
conclusions about storage capacity based on small networks alone, \ but it
is nonetheless instructive to make some comparative studies of pattern
stability in small networks.\ \ For several small values of $N$ and $p$, we
generated random sets of stored patterns and tested their stability, using
the HN and the BGN at several different values of $\gamma$. \ \ We counted
the average fraction of memory patterns which were stable, $f_{stable}$.%
\footnote{%
\ In the thermodynamic limit there is an approximate permutation symmetry
among the memory patterns, so that in general either all will be stable or
none will be. \ By contrast, in small networks it is not unusual for some
memorized patterns to reamain stable while others are not.} \ \ In addition,
we estimated the fraction of configuration space $V_{spur}$ occupied by
spurious attractors by following the trajectories of random initial
conditions inside the hypercube $\left\vert x_{i}\right\vert <2.$ These
results are collected in table \ref{smallnettable}. \ \ In all cases the
results were averaged over at least 100 sets of randomly generated patterns.
\ In generating random sets of patterns, we did not eliminate cases in which
two or more patterns are identical. \ \ The results show that, contrary to
the selected examples discussed in \cite{Chinarov} and \cite{IWANN}, the
spurious attractors do \emph{not} generically occupy much less configuration
space in the BGN than in the HN (in fact they normally occupy more,
especially at low values of $\gamma$). \ However, \ for the most part the
percentage of memory patterns which are stable is larger in the BGN than in
the HN as long as $\gamma\lesssim1$. \ As with larger networks, the BGN
becomes most similar to the HN when $\gamma>1,$ while lower $\gamma$ leads
to increased pattern stability. \ \ \ The increased pattern stability at low 
$\gamma$ is associated, however, with smaller basins of attraction for the
memory states and therefore with a greater volume of phase space occupied by
spurious states. \ 

\begin{table}[tbp] \centering%
\begin{tabular}{||l||l|l||l|l||l|l||l|l||}
\hline\hline
& \multicolumn{2}{l||}{$N=5,p=5$} & \multicolumn{2}{l||}{$N=10,p=5$} & 
\multicolumn{2}{l||}{$N=10,p=10$} & \multicolumn{2}{l||}{$N=15,p=10$} \\ 
\hline
$\alpha$ & $V_{spur}$. & $f_{stable}$ & $V_{spur}$ & $f_{stable}$ & $%
V_{spur} $ & $f_{stable}$ & $V_{spur}$. & $f_{stable}$ \\ \hline
0.25 & .38 & .87 & .96 & 1.00 & .85 & .92 & .97 & .97 \\ 
0.5 & .10 & .68 & .64 & .91 & .61 & .51 & .86 & .65 \\ 
0.75 & .06 & .49 & .51 & .75 & .56 & .32 & .83 & .37 \\ 
1.0 & .05 & .37 & .49 & .62 & .53 & .23 & .82 & .23 \\ 
1.5 & .08 & .37 & .51 & .47 & .53 & .16 & .80 & .15 \\ 
2 & 06 & .34 & .43 & .40 & .51 & .15 & .85 & .11 \\ 
(HN) & .03 & .34 & .38 & .48 & .50 & .16 & .75 & .15 \\ \hline\hline
& \multicolumn{2}{l||}{$N=20,p=5$} & \multicolumn{2}{l||}{$N=20,p=10$} & 
\multicolumn{2}{l||}{$N=50,p=5$} & \multicolumn{2}{l||}{$N=100,p=5$} \\ 
\hline
$\alpha$ & $V_{spur}$ & $f_{stable}$ & $V_{spur}$. & $f_{stable}$ & $%
V_{spur} $ & $f_{stable}$ & $V_{spur}$ & $f_{stable}$ \\ \hline
0.25 & 1.00 & 1.00 & 1.00 & .99 & 1.00 & 1.00 & 1.00 & 1.00 \\ 
0.5 & .92 & .99 & .95 & .74 & .99 & 1.00 & 1.00 & 1.00 \\ 
0.75 & .78 & .93 & .92 & .44 & .78 & 1.00 & .84 & 1.00 \\ 
1.0 & .71 & .86 & .92 & .26 & .61 & 1.00 & .55 & 1.00 \\ 
1.5 & .71 & .71 & .91 & .16 & .57 & .98 & .46 & 1.00 \\ 
2 & .74 & .62 & .93 & .11 & .57 & .97 & .46 & 1.00 \\ 
(HN) & .47 & .78 & .91 & .20 & .34 & 1.00 & .34 & 1.00 \\ \hline\hline
\end{tabular}
\caption{Properties of small networks\label{smallnettable}}%
\end{table}%

\section{\protect\bigskip Discussion\label{discussion}}

We have studied the properties of the BGN at high memory loading $\alpha$. \
Our results can be summarized as follows: \ For high values of $\gamma$,
such as 2, \ there is a first-order transition similar to that of a HN. \ \
For $\gamma=2$, the transition \ occurs at a critical loading $\alpha
_{c}\approx0.1$, which is lower than the critical loading of $\alpha
\approx0.148$ for the HN. \ As $\gamma$ decreases, the critical loading
increases and the phase transition evidently becomes weaker and eventually
disappears. \ For $\gamma=1$ the critical loading is apparently $\alpha
_{c}\approx0.17$ \ (higher than for the HN) and the phase transition is much
less pronounced for the finite-size networks we have studied. \ For $%
\gamma=0.5$ there is no evidence of a phase transition at all and patterns
are stable with very few errors up to $\alpha\approx0.3$. \ 

A phenomenon that occurs in the BGN much more than in the HN is the partial
retrieval of a pattern, whereby the dynamics corrects some sign errors in a
pattern without correcting all of them. \ This is especially noticeable in
the case of low $\gamma$ and high $\alpha$. \ This phenomenon suggests that
in this case the energy landscape in the vicinity of a stored pattern has
the shape of a funnel rather than a smooth basin of attraction. \ By a
smooth basin of attraction, we mean a connected neighborhood which is sloped
toward an attractor and in which there are few local minima that might
obstruct a trajectory once it has begun to flow toward the attractor. \ By a
funnel we mean a region with an overall average slope toward an attractor
which however contains many other local minima in which the trajectory might
become stuck before reaching the bottom. \ The presence of these local
minima is due to the bistability of the individual nodes and the local
potential barriers against spin flips. \ These same potential barriers are
also responsible for stabilizing the patterns. \ They make it less likely
that crosstalk noise will induce an error in a pattern, but they also can
inhibit the correction of an error which is present initially. \
Funnel-shaped landscapes were first examined in the context of protein
folding dynamics\cite{funnel1}\cite{funnel2}. \ It was suggested that at a
finite temperature such a landscape would allow the protein dynamics
efficiently to find the global minimum of energy in spite of the presence of
numerous local minima separated by potential barriers. If indeed the
landscape of the BGN at low $\gamma$ forms a funnel, then it is possible
that the introduction of some stochastic noise (i.e., finite temperature) \
could improve pattern retrieval by allowing trajectories to jump over the
comparatively small potential barriers into lower energy minima, just as in
the protein case. \ The effect of stochastic noise could be a fruitful
subject for further study. \ An interesting question is whether there is an
optimum level of noise which would improve the pattern retrieval ability of
the BGN with low $\gamma$ while at the same time maintaining the larger
storage capacity. \ \ Intriguing questions remain concerning the dynamics of
the BGN at low $\gamma$. \ The apparent initial expansion of the basins of
attraction of the memory states with increasing loading is counterintuitive,
and the patterns visible in figure \ref{scat05} hint at some dynamics which
is not yet fully understood. \ 

\begin{acknowledgments}
\bigskip This work was supported by Materials and Manufacturing Ontario
(MMO), a provincial centre of excellence. Discussions with V. Chinarov are
also acknowledged. \ 
\end{acknowledgments}

\bigskip


\begin{thebibliography}{99}
\bibitem{Chinarov} V. Chinarov and M. Menzinger, \ Biosystems \textbf{55},
137 (2000). \ 

\bibitem{LowLoading} P. McGraw and M. Menzinger, \ preprint
cond-mat/0203568, submitted to Phys. Rev. E. \ 

\bibitem{Hopfield} J.J. Hopfield, \ Proc. Natl. Acad. Sci. USA \textbf{79},
\ 2554 (1982).

\bibitem{Little} W.A. Little, Math. Biosci. \textbf{19}, 101 (1974). \ W.A.
Little and G.L. Shaw, Math. Biosci. \textbf{39}, 281 (1978).

\bibitem{Haykin} S. Haykin, Neural Networks: A Comprehensive Foundation. \
Prentice Hall, Upper Saddle River, NJ, 1999. \ 

\bibitem{Amitbook} Daniel J. Amit, \ Modeling Brain Function: \ The World Of
Attractor Neural Networks. \ Cambridge University Press 1989. \ 

\bibitem{HKP} J. Hertz, A. Krogh and R.G. Palmer, Introduction to the Theory
of Neural Computation. \ Perseus Books, Cambridge, MA 1991.

\bibitem{Amit1} D.J. Amit, H. Gutfreund and H. Sompolinsky, Phys.Rev. 
\textbf{A32}, 416 (1985); Phys. Rev. Lett. \textbf{55}, 428 (1985); \ 

\bibitem{Amit2} D.J. Amit, H. Gutfreund and H. Sompolinsky, \ Ann. Phys. 
\textbf{173}, 30 (1987). \ A. Chrisanti, D.J. Amit and H. Gutfreund,
Europhys. Lett. \textbf{2}, 337 (1986).

\bibitem{Hebb} D.O. Hebb, The Organization of Behavior, Wiley, New York,
1949. \ 

\bibitem{ContinuousHop} J.J. Hopfield, Proc. Natl. Acad. Sci. USA \textbf{81}%
, 3088 (1984).

\bibitem{IWANN} V. Chinarov and M. Menzinger, \ \textquotedblleft Bistable
Gradient Networks: Their Computational Properties\textquotedblright\ \ in J.
Mira and A. Prieto, eds., Connectionist Models of Neurons, Learning
Processes, and Artificial Intelligence, p.333. \ Springer Verlag, Berlin
2001.

\bibitem{Mezard} M. Mezard, G. Parisi and M. Virasoro, Spin Glass Theory and
Beyond. \ World Scientific, Singapore, 1987.

\bibitem{ChemNetE} W. Hohmann, M. Krauss and F.W. Schneider, J. Phys. Chem.
A \textbf{103}, 7606 (1999); \ J Phys. Chem A 102, 3103 (1998); \ J. Phys.
Chem A 101, 7364 (1997). \ G. Dechert, K.-P. Zeyer, D. Lebender and F.W.
Schneider, J. Phys Chem A 100, 19043 (1996).

\bibitem{ChemNetM} J.-P. Laplante, M. Pemberton, A. Hjelmfelt and J. Ross,
J. Phys. Chem 99, 10063 (1995). \ V. Booth, T. Erneux and J.-P. Laplante, J.
Phys. Chem 98, 6537 (1994). \ A. Hjelmfelt and J. Ross, J. Phys. Chem. 97,
7998 (1993). \ 

\bibitem{funnel1} P.E. Leopold, M. Montal and J.N. Onuchic, Proc. Natl.
Acad. Sci. USA \textbf{89}, 8721 (1992).

\bibitem{funnel2} P. Wolynes et al. Science \textbf{267}, 1618 (1995).

\bibitem{Karplus} O.M. Becker and M. Karplus, J. Chem Phys. \textbf{106},
1495 (1997).
\end{thebibliography}
\end{document}